\documentclass[12pt]{article}
\usepackage{graphicx}
\usepackage{amssymb}
\newcommand{\newc}{\newcommand}
\newc{\eeq}{\end{equation}}
\newc{\beq}{\begin{equation}}
\newc{\eeqa}{\end{eqnarray}}
\newc{\beqa}{\begin{eqnarray}}
\newc{\nonr}{\nonumber}
\newc{\bi}{\begin{itemize}}
\newc{\ei}{\end{itemize}}
\newc{\ra}{\rightarrow}
\newc{\LH}{\hat{L}}
\newc{\RH}{\hat{R}}
\begin{document}
\begin{titlepage}

\vskip 1.2cm

\begin{center}

{\Large \bf
Charged Lepton Electric Dipole Moments from TeV Scale Right-handed Neutrinos
}

\vskip 0.9cm

{\large
We-Fu Chang$^{a,b}$ and John N. Ng$^{a}$
}

\vskip 0.5cm

$^a$ {\it TRIUMF Theory Group,
           4004 Wesbrook Mall, Vancouver, BC V6T 2A3, Canada} \\
$^b$ {\it Institute of Physics, Academia Sinica,
         Taipei 115,  Taiwan} \\

\vskip 1.2cm

\abstract{ We study the connection between charged lepton electric
dipole moments, $d_l$ $(l=e,\mu,\tau)$,  and seesaw neutrino mass
generation in a simple two Higgs doublet extension of the Standard
Model plus three right-handed neutrinos (RHN) $N_a$, $a=1,2,3$. For RHN with
hierarchical masses and at least one with mass in the 10 TeV
range we obtain the upper bounds of $|d_e|< 9\times 10^{-30}$ e-cm
and $|d_{\mu}|<2 \times 10^{-26}$ e-cm. Our scenario favors the
normal mass hierarchy for the light neutrinos. We also calculated
the cross section for $e^-e^- \ra W^- W^-$ in a high luminosity
collider with constraints from neutrinoless double beta decay of
nuclei included. Among the rare muon decay experiments we find
that $\mu\ra e\gamma$ is most sensitive and the  upper limit is
$<8\times  10^{-13}$.

}

\end{center}
\end{titlepage}

\section{Introduction}
In the celebrated seesaw mechanism \cite{seesaw},
the right handed Majorana neutrinos are essential to generating small
Majorana masses for the active left-handed neutrinos of the Standard Model(SM).
These fields  are singlet under the SM gauge group and the exact number
required is open to debate. Since the light neutrino masses are constrained
to be less than an eV the masses of the right handed neutrinos have to be
heavier then $10^{12}$ GeV. This fits in well with expectations of
grand unified theory although the seesaw scale is lower
than the GUT scale which is generally taken to be around $10^{16}$ GeV as required by
proton stability. Moreover, the high scale also makes the seesaw mechanism impossible to test directly.
The best we can hope for are  indirect tests such as leptogenesis or renormalization effects. However,
in order to make predictions in these latter studies additional assumptions have to be made and the
results become highly model dependent.
On the other hand, Majorana masses for the light active neutrinos can be tested in neutrinoless
double beta decays of nuclei. Even in this case one has to eliminate other possible sources of
lepton number violating new physics such as exotic scalars. Recently there are attempts
to lower the masses of the right handed neutrinos to the TeV in leptogenesis studies \cite{lepto}. They
are particularly useful in supersymmetric models \cite{lows}. Clearly
such low scale Majorana neutrinos are of phenomenological and theoretical interests in their
own rights. They can be detected in high energy colliders and due to their rich CP
properties effects in low energy experiments can also be searched for. The prominent example is
the electric dipole moment(EDM) of a charged leptons denoted by $d_l$ where $l=e,\mu,\tau$.
Already the  experimental limit on $|d_e|$ is an impressive $< 10^{-27}$e-cm \cite{de}
and will be further improved in new round of experiments.
In contrast, the limit on  $|d_{\mu} |< 10^{-19}$ e-cm is
much less stringent and dedicated experiments are now being proposed.

In this paper we investigate the contribution of TeV scale Majorana right handed neutrinos,
$N_R$, to $d_l$.
An immediate issue is to decide whether they are involved in generating active neutrino
masses. Naively one expects that if they do so then the  seesaw mechanism will restrict their
Yukawa couplings to be very small
and thereby making their contribution to $d_l$ be minuscule. Thus, they can play an important role
in the seesaw mechanism  only in a subtle manner.
Radiative effects on the
left-handed part of the seesaw mass matrix was calculated in \cite{Pilaftsis:1991ug}.
Previously it was pointed out \cite{CNN} small Yukawa couplings can be avoided
 for more than one $N_R$.
We shall display this in the context of a simple model which consist of the SM plus
at least 3 right handed Majorana neutrinos, $N_R$, and an additional Higgs doublet. In order not to be
confused with possible CP violating phases from the scalar potential we assume one Higgs doublet couples to
$l_R$ and the other to $N_R$. This is the natural flavor conserving extended two Higgs doublet studied in
\cite{GW} and  is also well known to be part of the minimal supersymmetric standard model.
We shall see that one $N_R$ can be arranged to be heavy and is responsible for the seesaw and the other
two can be much lighter. Furthermore their Yukawa couplings can be of order unity. We do
not attempt a detail fit to the neutrino mixing data which can be a  separate study but merely to demonstrate
the possibility of such a scenario. This
very simple set up also  gives rise to a nonvanishing $d_l$ at the 2-loop level via a
set of Feynman diagrams specific to Majorana fermions as pointed out by \cite{NN};
a mechanism which has been  checked by \cite{Pil} and \cite{ACP}. The latter also contains a detail discussion of
the 2-loop integrals. However, we will concentrate more on the structure of CP violation that
appears and will be satisfied with order of magnitude estimates. We will be able to
give a `natural' order of magnitude estimate of the upper limit on $d_l$ coming from the possible existence of
multi-TeV scale $N_R$.

One would wonder why do we add one more Higgs doublet in our construction.
With only one Higgs doublet the 2-loop contribution to $d_l$
from Majorana neutrinos is negligible \cite{NN,ACP}. This is because now only the active
neutrinos take part as they couple to  the W bosons. Then $d_l$ is proportional
to their mass squared differences which
are known to be small from neutrino oscillation data. Thus, in the SM extended to include
seesaw neutrino mass although $d_l$ happens at 2-loop it is still undetectably small. In
contrast the SM with massless neutrinos $d_l$ receives contribution at 4-loop or higher.
With more than one Higgs doublet the physics changes. The right handed neutrinos do not
decouple  as we shall see later. This model can also  serve as a prototype in
studying the  interplay between scalars and Majorana fermions in EDM's.

 In section 2 we describe in detail a  model with three $N_R$'s.
We show how it works to generate sub-eV neutrino masses with
 one of them having mass in the  10 TeV range and he others can be much higher.
For early pioneering work on Majorana neutrinos in gauge theories see \cite{Valle}.

Next we discuss the Majorana phases in the model and dive into the $d_l$ estimates. Since the
physics scale we are interested in is relatively low the expected renormalization group running
of the parameters are not very significant and we shall ignore them.
Another possible test of
this mechanism can be done using the reaction $e^- e^- \rightarrow W^-W^-$; in the event that
the right handed neutrinos are not kinematically accessible at LHC or the linear collider.
This possibility has not been examined before and  is discussed in section 4. No neutrino
double beta decays ($0\nu \beta \beta$) of nuclei are discussed here.
Section 5 examines  the possible tests
in the rare decays $\mu \rightarrow e \gamma$ and
$\mu\rightarrow 3e$ using the constraint we found previously.
Discussions of other low energy probes
of Majorana phases can be found in \cite{GKM}. Finally we give our conclusions.

\section{A Simple Model with Right Handed Neutrinos}

The model we study is the SM with two Higgs doublet denoted by $\phi_1$ and $\phi_2$ and 3 right handed
neutrinos $N_a$ with $a=1,2,3$. The terms in the total Lagrangian of interest to us are  given by
\beqa
{\cal L}&=&
\frac{g_2}{\sqrt{2}} \overline{\nu_{Li}}\gamma^\mu
e_i W_\mu^+ + H.c. \nonr\\
&+& y_e^{ij} \overline{L_{i}}  \phi_1 e_{R j}
+  \zeta'_{i a}\overline{L_{i}}\tilde{\phi_2} N_a   + H.c.  \nonr\\
&-& \frac12 (M_{a b} \overline{N_a } N^{\mathbf{c}}_b + H.c.)
-V(\phi_1,\phi_2) +\cdots
\label{Basic}
\eeqa
where $\tilde{\phi}_2=i\sigma_2 \phi^*_2$ and indices $i,j=1,2,3$.
The hypercharge of Higgs doublets are $Y_1=Y_2=1/2$. This is the simplest
of two Higgs doublet models (2HDM). The details of the scalar
potential $V(\phi_1,\phi_2)$ is not important for us and will not spell out here.
Note that a $Z_2$ symmetry can be applied to fields so that the lagrangian is
invariant under the transformation:
\beq
 L_i\ra L_i \,,  \phi_1 \ra \phi_1\,,\; \phi_2 \ra -\phi_2\,,\;  N_a\ra -N_a \,,\; e_R\ra
 e_R\;.
\eeq
In this example,
the right-handed singlets only  couple to $\phi_2$ which is the simplest
way to enforce natural flavor conservation.
Other assignments to accommodate flavon models can also be applied. However, such
details are not necessary for us.
All the fermions are in the weak eigenbasis.
 $M_{ab}$ is a complex symmetric $3\times 3$ mass matrix for the
right-handed singlets  and $a,b=1,2, 3$.
Without loss of  generality, $M_{ab}$
can be chosen to be $diag\{M_1,M_2,M_3\}$ where the eigenvalues $M_a$ can be made
real and positive.
The Yukawa couplings  $y^{ij}$ and $\zeta'_{i a}$ are complex but the
Higgs potential are taken to be real.
In doing so the only source of possible CP violation come from the Yukawa couplings.

It is worthwhile to examine the number of physical phases in this
class of models. The discussion is clearest by first going into
the charged lepton mass basis. In this basis we denote the Yukawa
coupling by $\zeta'\ra\zeta$. Now consider the general case with
$N$ right-handed singlets and $n$ left-handed active neutrinos
coupling as above.  The neutrino mass matrix is a $(n+N)\times
(n+N)$ matrix in a block matrix form:

\beq
\left( \begin{array}{cc} 0&v\zeta\\
                 v\zeta^T&  M
\end{array} \right )
\label{bigM}
\eeq
where  $v \zeta$ is a $(n\times N)$ Dirac mass matrix and $v$ is a generic  vacuum expectation
value of the Higgs fields. For instance, for the Higgs doublet model of Eq.(\ref{Basic})
the following substitution  should be made: $v \ra v\sin \beta$ where $\tan \beta
=v_2/v_1$. Returning to Eq.(\ref{bigM}),
the lower block matrix, ${ M}$,  is
the  $(N\times N)$ Majorana mass matrix of the $N_R$'s. It is symmetrical and complex
and thus contain  $N(N+1)/2$ phases.
These  can be completely absorbed by the $U(N)$ mixing among the number $N$ of
$N_R$ states. Another way to view this is to use the freedom of phase choice in  $N_R$
 to rotate away the $N$ complex phases in the  eigenvalues.
Once that is done the phases of $N_R$ are fixed. The phases of the active $\nu_L$ are
still free. A phase redefinition of the $\nu_L$  will then remove $n$ phases  from $(n\times N)$ Yukawa
terms. This leaves a total of $n(N-1)$ physical phases \cite{Santa}. For the case
of $n=3$ and $N=3$ we have 6 physical phases. It is customary to assign one
phase to the light neutrino mixing matrix and leave the others in the mass eigenvalues.
Moreover, the physics of EDM is seen more clearly using complex Yukawa couplings.

The light neutrino Majorana mass
matrix can be solved from Eq.(\ref{bigM}):

\beq m_{\nu}= - v^2\zeta  M^{-1} \zeta^T. \eeq Explicitly, the
matrix elements are \beq \label{seesaw} m_{\nu,ij}= - v^2\left(
\frac{\zeta_{i1} \zeta_{j1}}{M_1}
+\frac{\zeta_{i2}\zeta_{j2}}{M_2}
+\frac{\zeta_{i3}\zeta_{j3}}{M_3} \right) \eeq where
$i,j=e,\mu,\tau$. The standard seesaw mechanism is to assume
$M_1\sim M_2\sim M_3 \sim 10^{14}$ GeV and the Yukawa couplings
are all of order unity so as to get sub eV masses for the active
neutrinos.
{
It is also noted by many that it hard to obtain the
observed bilarge mixing of the active neutrinos with inverse hierarchical masses.
}
Without more
assumptions we can  extract one more result, i.e.
\beq |\det
m_{\nu}|=\frac{v^6(\det \zeta)^2}{M_1M_2M_3}\;.
\eeq
which may be useful for constructing neutrino mass models.

On closer examination of  Eq.(\ref{seesaw}) one discovers other ways of getting small neutrino masses.
First  we scale out the lowest of the three $N_R$ masses which we call $M_<$.
Thus, Eq.(\ref{seesaw}) becomes
\beq
m_{\nu,ij}=-\frac{v^2}{M_<}\left(\frac{\zeta_{i1}\zeta_{j1}}{r_1}+\frac{\zeta_{i2}\zeta_{j2}}{r_2}
+\frac{\zeta_{i3}\zeta_{j3}}{r_3}
\right)
\label{ss2}
\eeq
where $r_a\equiv M_a/M_<,\ a=1,2,3$ and $r_a\geq 1$ by construction.
Each term in Eq.(\ref{ss2}) can be
view as a complex vector and it is a sum of  three such vectors.
If they form a triangle than the
element vanishes. The smallness of the active neutrino masses can then be due to
nearly closing of the complex triangle even with a value of $M_<$ in the TeV range.
Similar techniques have been used to
construct different hierarchies for the $N_R$ to yield the experimentally acceptable mass matrices
for $m_{\nu}$ \cite{King}. We leave aside the question of whether this is fining tuning
or manifestation of approximate family symmetry of the heavy neutrinos. We take it to
be purely phenomenologically motivated.

As an example, we describe a scenario in which 3 $N_R$ can
generate sub-eV active neutrino mass but possesses the features that some of them
are light enough, say $\sim$ TeV,
so that the seesaw mechanism can be amenable to testing in the near future.
Assume that $\zeta_{ia}$ obey the following relation:
\beq
\label{tri}
{\zeta_{i1} \over \sqrt{r_1}} \left(1-\frac{\delta_\nu}{2}\right)
= { \zeta_{i2} \over \sqrt{r_2}} \exp(i \pi/3)
= { \zeta_{i3} \over \sqrt{r_3}}\exp(i 2\pi/3)
\eeq
where $\delta_{\nu}$ is a small parameter we introduced. Then the seesaw mass contribution to the
 lights
neutrino from the 3 $N_R$'s nearly cancel among themselves. In the limit $\delta_{\nu}=0$,
Eq.(\ref{tri}) has a geometrical interpretation, i.e. the three terms viewed as vectors in the
complex plane forms an equilateral triangle. Then a non-vanishing $\delta_{\nu}$ is a measure
of the deviation from this configuration. In our example
 the $\zeta_1$ side is $(1+\delta_\nu )$ longer than the sum of the
other two and the small part left is responsible for small value of
$m_{\nu}$. As noted before
the model has six physical phases and there are nine complex Yukawa couplings.  For definiteness we
will choose the three $\zeta_{i1}$s to be real. Clearly the deviation $\delta_{\nu}$
can be associated with either one of
the $M_a$ and our choice is for simplicity of discussion.
To obtain an acceptable mass matrix we need further assumptions. Taking a hint from the charged leptons
we assume the couplings $\zeta_{ea}$ are such that
 $\zeta_{e a}\ll \zeta_{\mu a}\sim \zeta_{\tau a} \cong \zeta_a$. Then
a light neutrino mass matrix of the normal hierarchy type emerges
\beq
m_\nu \sim {\delta_\nu \zeta_1^2 v^2 \over r_1 M_<} \left( \begin{array}{ccc}
  0&0&0\\ 0&1&1\\  0&1&1
\end{array}\right).
\label{normal}
\eeq
It is  well known that Eq.(\ref{normal}) gives
the observed bilarge mixing angles. Notice the scale of $m_{\nu}$ is set by $M_1$ and
the parameters  $\delta_{\nu}$ and $\zeta$ also play a crucial role in determining its
magnitude.
Moreover, it is  important to note  that  the Yukawa
couplings are complex as explicitly displayed in Eq.(\ref{tri}).
These are physical phases which will enter into EDM considerations.

Interestingly we can extract more information about possible hierarchies in $M_a$. There are three
cases we can imagine:
\renewcommand{\theenumi}{\Roman{enumi}}
\begin{enumerate}
\item
\label{n1}
$M_1 \sim 10^{12}$ GeV and it sets the scale for light active neutrinos. If we take $\delta_{\nu} \sim \zeta_1 \sim 0.1$ then this
is sufficient to ensure sub-eV neutrinos. As seen in Eq.(\ref{normal})
$M_2$ and $ M_3$ play no role in determining $m_{\nu}$; thus  they can be as light as a few TeV. However,
Eq.(\ref{tri}) dictates that $\zeta_{1,2}$ will be
very small and hence will not give a detectable $d_l$.
\item \label{n2}
$M_1 \ll M_2 \lesssim M_3$ with  $M_1 \sim 10$TeV. For light neutrinos masses in the sub-eV range
Eq.(\ref{normal}) requires
the product $\delta_{\nu}\zeta_1^2 \lesssim 10^{-9}$. Superficially one would expect that
$\delta_{\nu}\sim \zeta_1 \sim 10^{-3}$. Since we do not have a theoretical basis for
 the values of
these two parameters it is prudent to use experimental constraints. We shall see later  that
neutinoless double beta decays limit  $\zeta_1 <0.1$.
 From Eq.(\ref{tri}) we see that
even if $M_3$ is of order $10^8$ GeV the corresponding $\zeta$ will be of order unity. This
is the interesting case for $d_l$. Certainly we can lower $M_1$ by simultaneously reducing $\delta_{\nu}$
 or $\zeta_1$.
\item \label{n3}
$M_1\sim M_2 \sim M_3$ and they are in the TeV range. In this case,
 either all the Yukawa couplings are small
or GIM-like cancellation with small mass splittings will have to take place.
Either case no interesting
$d_l$ arises.
\end{enumerate}
The above discussion is sufficient to illustrate the connection between the seesaw mechanism and
$d_l$ we set forth to seek.  Now we move on to the discussion of EDM.
\section{EDMs of Charged Leptons }

\begin{figure}[tc]
  \centering
  \includegraphics[width=0.5\textwidth]{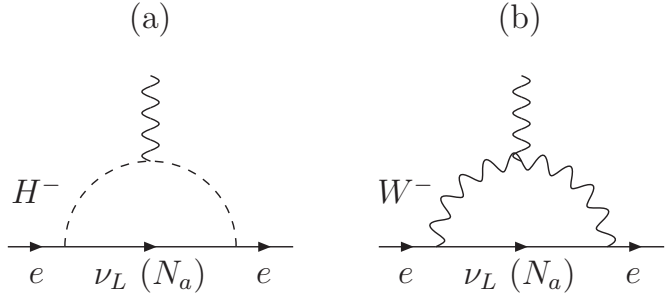}
  \caption{ 1-loop diagrams which may give no-zero EDM.}
\label{fig:1lp_EDM}
\end{figure}

In models with charged scalars and Majorana neutrinos, $d_l$ can begin at 1-loop.
The relevant Feynman diagrams are given in
Fig.\ref{fig:1lp_EDM}.
We shall employ the mass eigenstates in our discussions. The mixings between the light active
neutrinos and the heavy $N_R$'s are expected to be very small. Indeed from the
simple case of one family seesaw this mixing is given by $\zeta v/M$. For the parameter values
discuss in case (II) above we estimate  mixing between an active neutrino and the right-handed
singlet which we generically called $\theta$ to be  $\lesssim  10^{-3}$.

As seen in Fig.(\ref{fig:1lp_EDM}a)
both  heavy right-handed neutrinos and light active neutrinos can enter. Moreover, the Yukawa couplings are
conjugate of each other. Also the necessary helicity flip occurs in one of the external charged lepton lines; hence,
 there is no EDM from this diagram.
For the W boson exchange diagram, see Fig.\ref{fig:1lp_EDM}(b),
 active neutrinos exchanges are dominant with a small admixture of the right-handed neutrinos entering.
In either case the two $W$ vertices are also conjugate to each
other and clearly there is no EDM from this diagram.

\begin{figure}[h]
  \centering
  \includegraphics[width=0.8\textwidth]{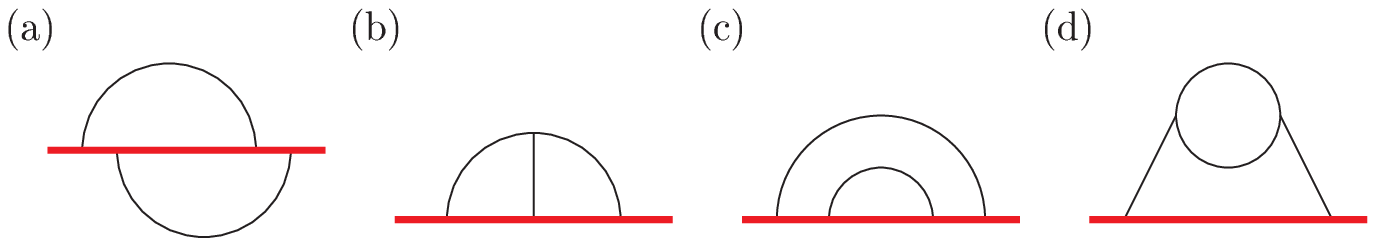}
  \caption{The topology of 2 loop diagrams, the thick lines represent
  the fermion lines, and the thin ones could be scalars, gauge boson,
  or fermion loop(sub-diagram d). }
\label{fig:2lp_TOP}
\end{figure}

At the two loop level there are 4 distinct topology we have to consider as shown in
Fig.\ref{fig:2lp_TOP}.
Fig.\ref{fig:2lp_TOP}(b-d) do not give rise to $d_l$ when the thin lines represent
gauge bosons. This  is well known form the SM.
When they represent scalar particles the Yukawa couplings
involve come in conjugate pairs ; thus negating their contributions.
Since we have no phases in the scalar sector we arrive at the result that
 Fig.\ref{fig:2lp_TOP}(b-d) give no EDM.
This leaves only Fig.\ref{fig:2lp_TOP}(a) as the only type that can lead to a non-vanishing $d_l$.

\begin{figure}[h]
  \centering
\includegraphics[width=0.8\textwidth]{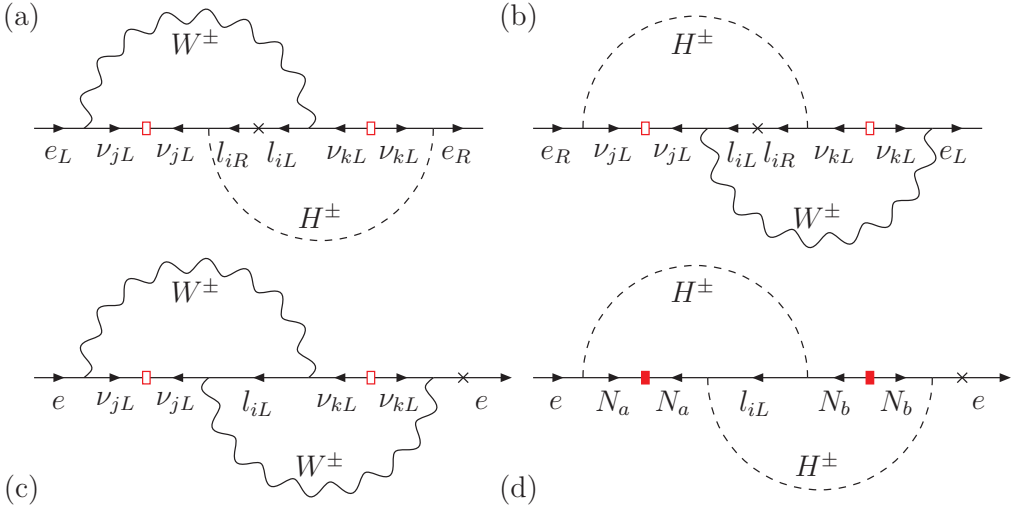}
  \caption{ 2-loop diagrams which give non-vanishing EDM. }
  \label{fig:2lp_EDM}
\end{figure}

To see the physics more clearly we put the details in Fig.\ref{fig:2lp_EDM}.
The external photon can attach to any charged object in the loops.
The Majorana mass insertions for the light neutrinos are indicated by the open box.
These also flip helicity
 and change lepton numbers by 2 units. The corresponding insertions for the heavy $N_R$'s
are denoted by the filled boxes. The diagrams with the internal $\nu_j $ lines replaced by $N_a$ or vice versa
are multiplied by the mixing $\theta$ alluded in the discussions at the beginning of this section and
will be suppressed.

It is instructive to examine the two W-boson diagram. It has two open box insertions which
involve light active neutrinos. Summations over different neutrino species and  the
three charged leptons $l_i$ are to be taken. The active neutrino mixing matrix elements at the incoming and outgoing lepton
vertices can be different; thus leading to a non-vanishing $d_l$. The open box
insertions indicating the lepton number violating nature of the Majorana masses are mandatory
 for this to happen. They do not exist for Dirac neutrinos and thus $d_l$ cannot happen at the
two loop level with Dirac neutrinos. Explicit calculations \cite{NN,ACP}show  that this  diagram gives a
contribution  proportional to $m_{\nu}^2$ and hence
is completely negligible. Similarly, the diagrams Fig.\ref{fig:2lp_EDM}(a) and (b) are suppressed by
 powers
of $m_{\nu}/M_{\mathrm W}$ and also $y_e \sim 10^{-6}$ . These graphs can be neglected.
This leaves only the two charged Higgs
exchange diagram.
The lepton EDM can be estimated as:
\beq
\frac{d_l}{e}
\sim \sum_{a<b} {m_l \over (16 \pi^2)^2}\sum_{i=e,\mu,\tau} Im[\zeta^*_{l a}\zeta_{l b}\zeta^*_{i a}\zeta_{i b}]
{M_a-M_b \over(M_a+M_b)^3}\ln \frac{M_H}{(M_a+M_b)}
\label{edmresult}
\eeq
in the limit that the $N_R$ are heavier that the charged Higgs boson which we assume to be of the weak
scale.
Strictly speaking, in the mass eigenbasis, the Yukawa couplings should be modified due to the
 mixing with the active neutrinos.
These are expected to be small, i.e. ${\cal O}(m_{\nu}/M)$, and can be neglected.
Note that the imaginary part of the product of four Yukawa couplings
flips sign  when one exchanges indices $a\leftrightarrow b$.
 In other words, only the antisymmetric
part in the loop integral yields the desired EDM operator.
On the other hand, the factor $(M_a-M_b)$ also reflects that the CP violating
effects go away when the masses of two right-handed  neutrinos become degenerate.
It is  also interesting to note that the diagram with the  photon attached to the
internal charged lepton has no EDM contribution since the loop integral
is completely  symmetric in $a$ and $b$.

From Eq.(\ref{edmresult}) we see that the EDM scales linearly as the mass of the charged lepton.
For the hierarchical mass of case (II) and take $M_2 \sim M_3 \sim 10^{8}$ GeV and $M_H \sim 200$ GeV
we obtain for the electron EDM
\beq
\label{de}
|d_e| \sim 9.2\times 10^{-31}(10 \mathrm{TeV}/M_1)^2 |\zeta_{e1}/0.1|^2 |\zeta_{1}|^2
\; \makebox{ e-cm}\;.
\eeq
The above estimate is not very sensitive to the values of $M_H$ and $M_{2,3}$ since their dependence is
 logarithmic.
Notice that we use a  small value of $\zeta_{e1}$ as is required by the normal hierarchy solution
and $0\nu \beta \beta$ (see next section). If the
right-handed neutrino masses are hierarchical such as $M_1<< M_2 <M_3$ then we have
\beq
\label{deh}
|d_e| \sim 9.2\times 10^{-31}(10 \mathrm{TeV}/M_1)^2 (M_2/M_3) |\zeta_{e1}/0.1|^2 |\zeta_{1}|^2 \; \makebox{ e-cm}
\eeq
and will be suppressed compared to Eq.(\ref{de}). Our estimate of $d_e$ is three orders of magnitude below
current experimental limit \cite{eedm} but is within reach of new  plan experiments \cite{edmmol}. We
note that Eq.(\ref{edmresult}) is a good approximation to the actual two Feynman integrals which cannot
be given in analytical form. Our numerical investigations show that it is accurate for order of magnitude
estimates.

We can also give an estimate of the muon EDM and it is
\beq
\label{muedm}
|d_{\mu}| \sim 1.8 \times 10^{-26} (10 \mathrm{TeV}/M_1)^2 |\zeta_1|^4 \; \makebox{ e-cm}\;.
\eeq
In this case the internal $\tau $ and $\mu$ lines give important contributions.
When combined the coupling involves is $\zeta_1$ which need not be small.
In contrast  $\zeta_{e1}$ which is small enters the calculation
for $d_e$ ( see Eq.(\ref{deh}) ).
Thus it is possible that $d_{\mu}$ can be more enhanced than just the mass factor
$m_\mu /m_e$ when compared to $d_e$.
The earlier discussion on the mass scaling violation of muon EDM
is given in \cite{Mscal}.
This illustrates the importance of doing both types of measurements. We add that our estimate is
six to seven orders of magnitude lower than current limit \cite{muedm} and will be a challenge
even for the newly  proposed dedicated $d_{\mu}$ measurements \cite{newmu}.

\section{ $(0\nu\beta\beta)$ decay and its inverse}

Here we discuss how low scale $N_R$ affects the decay rates of
$(0\nu\beta\beta)$ decays of nuclei. We will be concerned with the
elementary quark level $dd\rightarrow eeuu$ transition, and not
worry about the detail nuclear physics. At the fundamental fermion
level the amplitude is given by the diagrams below
\begin{figure}[h]
  \centering
\includegraphics[width=0.8\textwidth]{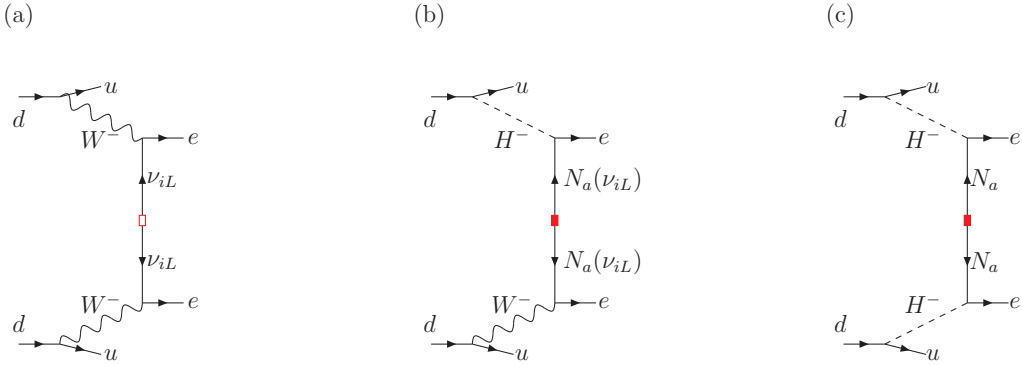}
  \caption{Tree level $(0\nu \beta \beta)$ decay amplitudes from Majorana neutrinos}
  \label{fig:0nu}
\end{figure}

An estimate of the amplitude for the 2W exchange diagram is
\beq
A_a \sim {g^4}\frac{1}{M_W^4}\frac {m_{\nu,ee}}{ \langle p \rangle^2}
\eeq
where $\langle p \rangle$ is the average momentum of the exchange light neutrino. The corresponding diagram
with $N$ replacing $\nu$ line is suppressed by the $M$ and $\theta^2$.
For the amplitude of Fig.\ref{fig:0nu}(b) with $\nu$ exchange we estimate
\beq
A_b\sim g^2 {m_q m_e\over M_W^2}{1\over M_W^2 M_H^2}{m_{\nu, ee}\over
\langle p \rangle^2}\sim 3\times 10^{-11} A_a\; ,
\eeq
where $m_q$ represents the light quark mass,
and with $N$ exchange we obtain
\beq
A_b\sim g^2 {m_q \over M_W}{\zeta \over M_W^2 M_H^2}{1\over M_N}\theta 
\sim 3\times 10^{-4} A_a\;.
\eeq
Similarly the dominant $2H$ exchange graph is associated with a  $N$-line and it gives
\beq
A_c\sim g^2 {m_q^2 \over M_W^2}{\zeta^2 \over  M_H^4}{1\over M_N}
\sim  6\times 10^{-8} A_a\;.
\eeq
In arriving the above estimation, we have used the following numerical numbers:
$M_N\sim 10$ TeV, $M_W\sim 100$GeV, $M_H\sim 200$ GeV,
$m_{\nu,ee}\sim 10^{-10}$GeV, the quark mass $m_q \sim 10^{-3}$ GeV and $\langle p \rangle \sim 0.1 GeV$ \cite{pnu}
This analysis suggests that $WW$ exchange with active Majorana neutrino  exchange  is still the dominated tree-level contribution
to $(0\nu\beta\beta)$ decay.

Interestingly, for low scale right-handed Majorana neutrino
 important contribution to $(0\nu \beta\beta)$ decays can come from 1-loop diagrams depicted below
\begin{figure}[htc]
  \centering
  \includegraphics[width=0.5\textwidth]{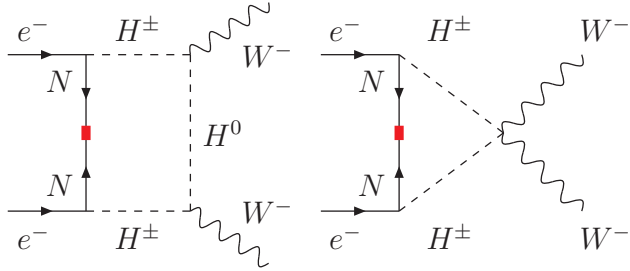}
  \caption{The box and triangle diagrams for $(0\nu\beta\beta)$ decay. }
  \label{fig:1lp_eeWW}
\end{figure}

The effective Lagrangian the diagrams generate is
\beqa
\label{eeww}
{\cal{L}}&=& F_l \;\;\left( \bar{e^c}\LH e\right) \epsilon_1\cdot \epsilon_2\,, \nonr \\
F_l&=&\frac{c g^2}{32\pi^2}\sum_a \frac{\zeta_{ea}^2}{M_a}\ln \left(\frac{M_H}{ M_a}\right)\,
\eeqa
where $c$ is an order one constant which depends on the details of the 2HDM such as scalar
mixings.
The $\epsilon_{1,2}$  are polarization 4-vectors of the $W$ bosons. We have also assumed  $M_a \gg M_H$ with
$M_H$ denoting a common scalar mass.
After dropping the  $m_{\nu, ee}$ part the sum in the above equation yields
\beq
\label{form}
F_l=\frac{ \alpha }{8\pi \sin^2\theta_W}  \frac{\zeta_{e1}^2}{M_1}
\left[\ln r_2 e^{-\frac{2\pi i}{3}} +\ln r_3 e^{-\frac{4\pi i}{3}}\right]\,
\eeq
by using Eq.(\ref{tri}) and assuming $c=1$ for simplicity.
If this is the dominant contribution to $(0\nu\beta\beta)$ decays then
we can set the limit $\zeta_{e1} <0.1$ for $M_1= 10 \mathrm {TeV}$. In comparison the low
energy effective Lagrangian from the exchange of $N_1$ is given by
\beqa
{\cal {L}}_{tree}&=& F_t \;\;\left( \bar{e^c}\gamma_{\mu}\gamma_{\nu}\LH e \right)
\epsilon_1^{\mu} \epsilon_2^{\nu}\,,\nonr \\
F_t&=& \frac{4\pi \alpha}{\sin^2\theta_W}\left(\frac{\theta^2}{M_1}\right)\,.
\eeqa
The numerical absolute value of the square bracket in Eq.(\ref{form}) is $\sim 5$ for a large range
of $r_2$ and $r_3$. For $\zeta_{e1}=.1$ and $M_1= 10 {\mathrm {TeV}}$ we get
$F_l=6\times 10^{-9}$ whereas  $F_t= 4\times 10^{-11}$ in units of ${\mathrm {GeV}}^{-1}$.
Thus, the loop diagram can be more important even when $\zeta_{e1}$ is not that large.

The effective Lagrangian of Eq.(\ref{eeww}) can also give rise to 2 W-boson production in $e^- e^-$
colliders even when $N_1$ is too heavy to be directly produced. The cross section can be
easily calculated to be
\beq
\label{sig}
\sigma (ee\rightarrow WW) = \frac{F_l^2}{128\pi}\frac{s^2}{M_W^4}
\eeq
which is dominated by the longitudinal components of the $W$ bosons. Although we have neglected
the energy dependence in $F_l$ this is sufficient for a ball park estimate of the cross
section. For a linear collider with $\sqrt s = 2$ TeV and a luminosity of
$10^{34}{\mathrm {sec}}^{-1}{\mathrm {cm}}^{-2}$ we obtain
1.3 events in a 100 days running for a 10 TeV Majorana neutrinos. As mandated by the transient
high $s$ behavior of Eq.(\ref{sig}) one would require the highest available energy
for  a given high luminosity collider to probe this physics.
It is easy to check that the usual tree level $t$-channel $N_R$ exchange mechanism in the seesaw
scenario gives a even smaller cross section \cite{ee2W}.

\section{ A Trio of Rare Muon Decays}
\begin{figure}[htc]
  \centering
  \includegraphics[width=0.2\textwidth]{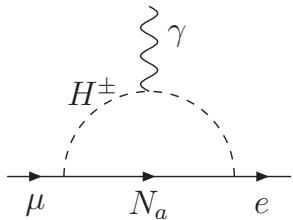}
  \caption{ Right-handed Majorana neutrino contribution to  $\mu\ra e \gamma$ decay.}
  \label{fig:MuEA}
\end{figure}
The rare decays $\mu \ra e+\gamma, 3e $ and $\mu-e$ conversion in nuclei  have always been a favorite for testing models of
lepton violations. If TeV scale Majorana $N_R$'s  exist one  also expects that these processes will
 occur. A general up to date review is given in \cite{KO} and we shall follow the notations used there.

We begin with  $\mu\ra e \gamma$. The most general Lorentz and gauge invariant $\mu-e-\gamma$ interaction
is given by
\beqa
{\cal M}= -e A^*_\lambda \overline{u_e}(p_e)\left\{
\left[ f_{E0}(q^2)+f_{M0}(q^2) \gamma^5\right]\gamma_\nu
\left(g^{\lambda\nu}-{q^\lambda q^\nu \over q^2} \right)\right.\nonr\\
\left. + \left[ f_{M1}(q^2)+f_{E1}(q^2) \gamma^5\right]{i\sigma^{\lambda\nu}q_\nu \over m_\mu}
\right\}u_\mu(p_\mu)
\eeqa
where $q^{\lambda }$ and $A_{\lambda}$ are the photon 4-momentum and polarization respectively and
$p_e=p_{\mu}-q$. For $\mu \ra e\gamma$ only the form factors $f_{M1}$ and $f_{E1}$
contribute. They can be calculated from
 the dominant diagram given by Fig.(\ref{fig:MuEA}).
 The transition rate is
\beq
\label{MueA}
B(\mu \ra e \gamma)=
 {3 \alpha\over 64 \pi}{ |\zeta_{e 1}\zeta^*_{\mu 1}|^2 \over G_F^2 M_1^4}
= 8.01\times 10^{-11} |\zeta_{e 1}\zeta^*_{\mu 1}|^2
\left( {10 \mbox{TeV} \over M_1}\right)^4\;.
\eeq
In arriving at the last formula, we have ignored the small $M_H/M_a$ term. The current
experimental limit of $< 1.2 \times 10^{-11}$ \cite{PDG} sets a loose constrain on the mixings.
Alternatively we can take $\zeta_{e1}<0.1$ as required by our model and $\zeta_{\mu 1}<1$
so as not to have strong Yukawa; then we get an upper limit of $8\times 10^{-13}$ for
a 10 TeV $N_R$.
We note in passing similar decays for the $\tau$ are sensitive to the mixing $\zeta_{\tau 1}$
which will be hard to obtain from other experiments.

For $\mu-e$ conversion in nuclei the seesaw model belongs to the class where the photonic
penguin diagram as given Fig.\ref{fig:MuEA} with the photon off shell dominates the transition
rate. An explicit calculation gives
\beq
B_{\rm conv} = { m_\mu^5 G_F^2 F_p^2 \alpha^4 Z_{eff}^4 Z \over 12\pi^3\Gamma_{\rm capt} }
B(\mu\ra e+\gamma)\;.
\eeq
For $\,^{48}_{22}Ti$, $F_p=0.55$, $Z_{eff}=17.61$ and $\Gamma_{\rm capt}=1.71\times 10^{-18}
GeV$ which implies $B^{Ti}_{\rm conv}\sim 0.004 B(\mu\ra e+\gamma)$.

Similarly for $\mu\ra 3e$ the photonic penguin is the most  important graph.
The box diagram with two charged Higgs exchange is completely negligible. The Z-penguin
graphs are also sub-dominant. Thus we obtain simply

\beq
B(\mu\ra 3e) =
 {2\alpha\over 3\pi}\left(\ln\frac{m_\mu}{m_e}-\frac{11}{8}\right)
B(\mu\ra e \gamma)
\eeq
or $B(\mu\ra 3e) \sim 0.006 B(\mu\ra e \gamma) $.
Hence, $\mu\ra e \gamma$ and $\tau \ra \mu(e) \gamma$ are the most
important processes to probe TeV scale seesaw.

\section {Conclusions}

In this paper we have given a detailed study of the connection between seesaw neutrino mass generation and
charged lepton EDMs. The  2HDM we employed is simple and it captures the physics clearly and succinctly.
It is expected to be a crucial part of any elaborate embedding of the seesaw mechanism into a
grand unified picture. As noted previously if all the right-handed neutrinos  have very
high masses, i.e. $> 10^{10}$ GeV then $d_l$ will be undetectably small.
We found that it is crucial to have at least one $N_R$ have a mass in  10 TeV or slightly lower range.
In addition, not all the Yukawa couplings can be suppressed as in the charged leptons.
The charged lepton EDM arises from two loop diagrams involving Majorana neutrinos and
the associated physical phases
that have no counter parts in the SM with Dirac neutrinos. Under favorable choice of parameters
we estimated that the upper limits  are  $|d_e| < 9\times 10^{-30}$e-cm  and $|d_{\mu}|< 1.8 \times 10^{-26}$
e-cm for a 10 TeV Majorana neutrino. The parameters involved are consistent with a normal
mass hierarchy for the light active neutrinos.

Interestingly, a right-handed neutrino in the 10 TeV mass range may also be
required for a successful leptogensis \cite{lows}.
It is reasonable to expect that the mass
 of the lowest Majorana neutrino is in the 10 TeV range especially in the supersymmetry context. Moreover,
the direct production of $N_1$ is out of reach for high energy colliders under discussion. Even so
we considered how a high luminosity $e^-e^-$ collider in the TeV range can still probe their existence via the
same sign $2W^-$ production. Complementing this we calculated that  the rare muon decay $\mu\ra e\gamma$
at the level of $10^{-13}$
is  found to be sensitive test of the  scenario we discussed. Similarly, one can contemplate the
rare Z decays into $\tau \mu$ or $\tau e$. The signatures are  clean and unmistakable. At 1-loop this
proceeds via a similar diagram as given in Fig.(\ref{fig:MuEA}). We estimate this to give
a branching ratio of $< 10^{-16}$ which is too small even for a Z-factory.

Interestingly, the CP violation from the see-saw mechanism has
negligible  effect in the quark sector. This demonstrates clearly that the search of neutron and
electron and muon  EDM's  are independent powerful probes of physics beyond the SM.

\section{Acknowledgement}
WFC wants to thank K. Cheung and C.Q. Geng for useful discussions.
This work is supported in part by the Natural Science and Engineering Research Council of Canada.

\end{document}